\documentclass[namedreferences]{solarphysics}
\usepackage[optionalrh]{spr-sola-addons} 
\usepackage{graphicx}
\usepackage{color}           
\usepackage{url}             
\usepackage{rotating}
\usepackage{epstopdf}

\newcommand{\jasr}{    {\it Adv. Space Res.}}

\newcommand{\aap}{    {\it Astron. Astrophys.}}

\newcommand{\apj}{    {\it Astrophys. J.}}
\newcommand{\apjl}{   {\it Astrophys. J. Lett.}}

\newcommand{\grl}{    {\it Geophys. Res. Lett.}}

\newcommand{\jastp}{  {\it J. Atmos. Solar-Terr. Phys.}}
\newcommand{\jgr}{    {\it J. Geophys. Res.}}
\newcommand{\jswsc}{    {\it Space Weather Space Clim.}}

\newcommand{\pasj}{   {\it Pub. Astron. Soc. Japan}}

\newcommand{\solphys}{{\it Solar Phys.}}

\newcommand{\ssr}{    {\it Space Sci. Rev.}}

\begin{document}

\begin{article}

\begin{opening}

\title{Statistical Evidence for Contributions of Flares and Coronal Mass Ejections to Major Solar Energetic Particle Events}

\author{G.~\surname{Trottet}$^{1}$\sep
          S.~\surname{Samwel}$^{2}$\sep
          K.-L.~\surname{Klein}$^{1}$\sep
          T.~\surname{Dudok de Wit}$^{3}$ \sep 
          R.~\surname{Miteva}$^{1, 4}$
       }
\runningauthor{G. Trottet et al.}
\runningtitle{Flare and CME Contributions to SEP Events}

   \institute{
             $^{1}$ Observatoire de Paris, LESIA-CNRS UMR 8109, Univ. P \& M Curie and Paris-Diderot, 
                                Observatoire de Meudon, F-92195 Meudon, France
                     email: \url{gerard.trottet@obspm.fr}, \url{ludwig.klein@obspm.fr}, \url{rositsa.miteva@obspm.fr} \\ 
              $^{2}$ National Research Institute of Astronomy and Geophysics (NRIAG), Helwan, Cairo, Egypt
                     email: \url{samwelsw@nriag.sci.eg} \\
               $^{3}$Laboratoire de Physique et Chimie de l'Environnement et de l'Espace, 3A, Avenue de la Recherche Scientifique, 45071 Orl\'eans cedex 2, France email: \url{ddwit@cnrs-orleans.fr} \\
               $^4$presently at Space Research and Technology Institute - Bulgarian Academy of Sciences, Acad. Georgy Bonchev st, 1113 Sofia, Bulgaria 
               }

\begin{abstract}
Solar energetic particle (SEP) events are related to flares and coronal mass ejections (CMEs). This work is a new investigation of statistical relationships between SEP peak intensities - deka-MeV protons and near-relativistic electrons - and characteristic quantities of the associated solar activity. We consider the speed of the CME and quantities describing the flare-related energy release: peak flux and fluence of soft X-ray (SXR) emission, fluence of microwave emission. The sample comprises 38 SEP events associated with strong SXR bursts (classes M and X) in the western solar hemisphere between 1997 and 2006, and where the flare-related particle acceleration is accompanied by radio bursts indicating electron escape to the interplanetary space. The main distinction of the present statistical analysis from earlier work is that besides the classical Pearson correlation coefficient the partial correlation coefficients are calculated in order to disentangle the effects of correlations between the solar parameters themselves. The classical correlation analysis shows the usual picture of correlations with broad scatter between SEP peak intensities and the different parameters of solar activity, and strong correlations between the solar activity parameters themselves. The partial correlation analysis shows that the only parameters that affect significantly the SEP intensity are the CME speed and the SXR fluence. The SXR peak flux and the microwave fluence have no additional contribution. We conclude that these findings bring statistical evidence that both flare acceleration and CME shock acceleration contribute to the deka-MeV proton and near-relativistic electron populations in large SEP events.
\end{abstract}


\end{opening}


\section{Introduction}\label{intro}

Transiently enhanced intensities of solar energetic particles (SEPs) in space are observed in association with flares and fast coronal mass ejections (CMEs). Potential accelerators exist in different regions of the corona: magnetic reconnection in the flaring active region and in the magnetically stressed corona in the aftermath of a CME, as well as the shock wave driven by a fast CME. Many attempts were made in the past to identify a unique  accelerator by establishing preferential statistical relationships between SEP parameters, especially their peak intensity, and one of the quantities describing the importance of the associated eruptive activity - flare radiation on the one hand, CME speed on the other \cite{Kah-01,Gop:al-03,Can:al-10,Mit:al-13,Ric:al-14}. Since measurements of soft X-ray (SXR) emission are readily available from the {\it Geostationary Operational Environmental Satellites} (GOES), they are frequently used to forecast and characterise the importance of an SEP event \cite{Gar-04,Kah:al-07a,Blc-08,Nun-11}. Comparisons of the pairwise correlation with SEP peak intensities has not been conclusive so far, since similar correlation coefficients were found for flare-related (SXR peak flux) and CME-related (speed) parameters, and a broad scatter. The interpretation is furthermore complicated by the fact that the solar parameters are not independent \cite{Kah-82b}.

A shortcoming of SXR emissions is that they reveal the state of the flare-heated plasma, not of flare-accelerated particles. The most direct radiative diagnostic of non thermal protons at the Sun is nuclear line emission in the 4-7~MeV energy range, notably due to the impact of protons of a few tens of MeV on the chromosphere.  \inlinecite{Chr-90} showed a close correlation between the gamma-ray fluence in this range and the peak proton intensity of SEP events at energies above 10~MeV. However, other authors presented evidence against such a relationship \cite{Cli:al-89,PEn:Mir-99,Ohk-03}. The small number of events with gamma-ray coverage leaves some uncertainty that can only be reduced by using proxies. One is radio emission at microwave frequencies, {\it i.e.} between 1~GHz and several tens of GHz. It is generated by the gyro-synchrotron process by near-relativistic electrons at energies between about 100~keV and a few MeV, and is routinely observed from ground. It was shown by \inlinecite{Ves-88}, \inlinecite{Mur:al-93}, \inlinecite{Chr-90} and \inlinecite{Shh:al-09} that nuclear gamma-ray line emission from protons at some tens of MeV and microwave or hard X-ray (HXR) emissions from electrons above 300 keV are well correlated. So one can use microwave emission as a proxy of both electrons and protons accelerated during flares. Statistical studies of the SEP-microwave relationship were published by \inlinecite{Kah-82b} and \citeauthor{Dbg:al-87} (\citeyear{Dbg:al-87}, \citeyear{Dbg:al-89}).

In the present work we carry out a comparative analysis of the correlation between SEP intensities - electrons between a few tens and a few hundreds of keV, protons at some tens of MeV - with quantities characterising the eruptive solar activity: CME speed, peak flux and fluence of the SXR emission and fluence of the microwave emission. The data sources and the event sample are presented in Section~\ref{Sec_Obs}. We identify those events where radio observations provide evidence that flare-accelerated particles escape to the interplanetary space (Section~\ref{Sec_Obs_esc}). Classical Pearson correlations between the intensities of electrons and protons detected in space, between the quantities of the associated eruptive activity, and between these quantities and the SEP peak intensities are evaluated in Sections \ref{Sec_corr_SESE} - \ref{Sec_corr_SoSE}. In Section~\ref{Sec_corr_part} partial correlations are used in the attempt to distinguish between correlations that suggest a physical relationship and those which are spuriously produced by the interdependence of different variables describing the solar activity. The results are summarised and discussed in Section~\ref{Sec_Disc} with respect to earlier work and to their implication on the sites and mechanisms of SEP acceleration.


\begin{sidewaystable}
\footnotesize
\caption[]
{Parameters of SoWi events: columns~(1) event date, (2) SXR start time, (3) quality flag for the microwave fluence calculation and particle escape, (4) and (5) frequency where the microwave fluence is maximum in GHz and peak fluence in 10$^5$sfu s (6) peak soft X-ray flux in the 0.1-0.8~nm  channel in 10$^{-4}$W m$^{-2}$, (7) start-to-peak fluence [10$^{-4}$J m$^{-2}$] in this wavelength range, (8) projected CME speed in km s$^{-1}$, (9)-(11) peak intensities in (cm$^2$ s sr MeV)$^{-1}$ of electrons (38-53 and 175-315~keV) and of protons (15-40~MeV).\\}
\vspace{1ex}
\label{Tab_SoWi}
\begin{tabular}{lrrrrrrrrrr}
\hline
\hline
\multicolumn{8}{c}{Solar Activity} & \multicolumn{3}{c}{SEP peak intensity} \\
Date             & Start & Qual & $\nu_{\rm max}$ & $\Phi_{\mu}$ & $I_{\rm SXR}$ & $\Phi_{\rm SXR}$ & $V_{\rm CME}$ & $J_{\rm e} (38 \; \rm keV)$ & $J_{\rm e}(175 \; \rm keV)$ & $J_{\rm p}(15 \; \rm MeV)$ \\
yyyy mm dd & hh:mm & & [GHz]    &              &           &              &           & ($\times 10^3$) & ($\times 10^3$) &  ($\times 1$) \\
(1) & (2) & (3) & (4) & (5) & (6) & (7) & (8) & (9) & (10) & (11) \\
\hline
1997 11 03  & 10:18 & 1 & 4.99 & 0.41 & 0.42 & 66.59 & 352 & 3.27 (16:00) & 0.04 (15:00) & 0.00 \\ 
2000 03 03  & 02:08 & 1c & 15.40 & 1.32 & 0.38 & 46.80 & 841 & 5.58 (03:00) & 0.18 (03:00) & 0.00 \\ 
2000 03 22  & 18:34 & 2c & 15.40 & 2.21 & 1.10 & 386.32 & 478 & 11.99 (20:45) & 0.06 (21:55) & 0.03 (22:35) \\ 
2000 04 04  & 15:12 & 1 & 2.69 & 3.45 & 0.10 & 105.30 & 1188 & 114.00 (16:00) & 0.79 (15:50) & 0.62 (33:25) \\ 
2000 06 15  & 19:38 & 1 & 4.99 & 0.63 & 0.18 & 114.25 & 1081 & 52.49 (20:15) & 0.30 (20:55) & 0.00 \\ 
2000 06 17  & 02:25 & 1c & 8.80 & 0.12 & 0.35 & 119.17 & 857 & 160.00 (06:45) & 0.86 (05:25) & 0.03 (07:50) \\ 
2000 11 24  & 04:55 & 1 & 17.00 & 11.43 & 2.00 & 230.08 & 1289 & 23.00 (08:10) & 0.78 (08:10) & 0.23 (11:40) \\ 
2000 11 24  & 14:51 & 2 & 15.40 & 19.40 & 2.30 & 734.04 & 1245 & 172.00 (17:54) & 5.44 (17:05) & 2.20 (18:09) \\ 
2001 03 10  & 04:00 & 1 & 17.00 & 0.52 & 0.67 & 60.64 & 819 & 8.17 (15:30) & 0.27 (09:25) & 0.00 \\ 
2001 04 10  & 05:06 & 1 & 8.80 & 68.14 & 2.30 & 1068.96 & 2411 & 61.20 (22:00) & 2.01 (17:15) & 2.88 (31:00) \\  
2001 04 26  & 11:26 & 1 & 8.80 & 1.36 & 0.78 & 0.00 & 1006 & 3.84 (15:45) & 0.03 (15:40) & 0.05 (42:05) \\ 
2001 10 19  & 00:47 & 1c & 8.80 & 6.36 & 1.60 & 547.39 & 558 & 8.18 (05:00) & 0.26 (04:55) & 0.19 (06:25)  \\ 
2001 10 19  & 16:13 & 1c & 8.80 & 14.25 & 1.60 & 654.63 & 901 & 18.60 19:00) & 0.33 (18:20) & 0.18 (19:35) \\ 
2001 11 04  & 16:03 & 1 & 4.99 & 36.93 & 1.00 & 406.09 & 1810 & 2590.00 (34:49) & 167.00 (36:15) & 147.00 (35:19) \\ 
2001 11 22  & 22:40 & 1 & 3.75 & 14.00 & 1.00 & 1300.00 & 1443 & 2570.00 (47:40) & 58.70 (41:35) & 177.00 (37:55) \\ 
2001 12 26  & 04:32 & 1 & 4.99 & 36.50 & 0.71 & 1188.53 & 1446 & 800.00 (07:15) & 25.50 (07:00) & 23.50 (11:10) \\ 
2002 02 20  & 05:52 & 1 & 17.00 & 1.14 & 0.51 & 100.05 & 952 & 375.00 (06:45) & 5.29 (06:30) & 0.55 (07:25)  \\ 
2002 07 15  & 19:59 & 1 & 8.80 & 14.97 & 3.00 & 455.24 & 1151 & 187.00 (49:30) & 5.34 (48:15) & 1.30 (47:15) \\ 
2002 08 14  & 01:47 & 1 & 2.69 & 1.78 & 0.23 & 211.94 & 1309 & 602.00 (02:30) & 3.34 (02:25) & 0.33 (13:10) \\ 
2002 08 24  & 00:49 & 1 & 9.40 & 95.86 & 3.10 & 1731.50 & 1913 & 156.00 (02:00) & 10.80 (02:25) & 10.06 (08:40) \\ 
2002 11 09  & 13:08 & 1 & 4.99 & 3.98 & 0.46 & 181.30 & 1838 & 99.10 (23:40) & 2.34 (23:40) & 3.53 (23:50)  \\ 
\hline
\end{tabular}
\end{sidewaystable}

\begin{sidewaystable}
\footnotesize
\addtocounter{table}{-1}
\caption[]
{Parameters of SoWi events (cont'd). \\}
\vspace{1ex}
\begin{tabular}{lrrrrrrrrrr}
\hline
\hline
\multicolumn{8}{c}{Solar Activity} & \multicolumn{3}{c}{SEP peak intensity} \\
Date & Time & Qual & $\nu_{\rm max}$ & $\Phi_{\mu}$ & $I_{\rm SXR}$ & $\Phi_{\rm SXR}$ & $V_{\rm CME}$ & $J_{\rm e} (38 \; \rm keV)$ & $J_{\rm e} (175 \; \rm keV)$ & $J_{\rm p}(15 \; \rm MeV)$ \\
yyyy mm dd & hh:mm & & [GHz]    &              &           &              &           & ($\times 10^3$) & ($\times 10^3$) &  ($\times 1$) \\
(1) & (2) & (3) & (4) & (5) & (6) & (7) & (8) & (9) & (10) & (11) \\
\hline
2003 03 18  & 11:51 & 1 & 4.99 & 28.50 & 1.50 & 448.12 & 1042 & 375.00 (13:20) & 2.98 (13:20) & 0.03 (16:35) \\ 
2003 04 23  & 00:56 & 1 & 9.40 &   3.26 & 0.51 & 104.55 &   916 &     0.87 (01:35) & 0.08 (02:35) &  0.03 (07:10) \\ 
2003 04 24  & 12:45 & 1 & 8.80 & 0.60 & 0.33 & 47.94 & 609 & 5.02 (13:30) & 0.05 (13:25) & 0.04 (15:30) \\ 
2004 04 11  & 03:54 & 2 & 2.69 & 1.58 & 0.10 & 61.64 & 1645 & 74.80 (12:45) & 1.43 (09:45) & 0.24 (10:05) \\ 
2004 07 13  & 00:09 & 1 & 17.00 & 1.91 & 0.67 & 94.64 & 607 & 1.12 (03:05) & 0.05 (02:50) & 0.03 (10:10) \\ 
2005 05 06  & 03:05 & 1 & 9.40 & 0.38 & 0.09 & 23.30 & 1120 & 75.70 (05:20) & 0.67 (04:24) & 0.05 (10:10) \\ 
2005 05 06  & 11:11 & 2 & 15.40 & 0.35 & 0.13 & 34.90 & 1144 & 78.60 (13:10) & 0.41 (12:35) & 0.04 (22:25) \\ 
2005 05 11  & 19:22 & 1 & 2.69 & 0.10 & 0.11 & 53.35 & 550 & 5.05 (22:20) & 0.05 (21:45) & 0.03 (25:25) \\ 
2005 07 13  & 14:01 & 1c & 4.99 & 51.40 & 0.50 & 741.17 & 1423 & 252.00 (17:09) & 2.28 (16:54) & 0.34 (26:50) \\ 
2005 08 22  & 00:44 & 1  & 4.99 & 34.00 & 0.26 & 423.81 & 1194 & 125.00 (05:25) & 1.40 (03:10) & 0.22 (06:30) \\ 
2005 08 22  & 16:46 & 1 & 4.99 & 212.44 & 0.56 & 666.10 & 2378 & 804.00 (23:05) & 12.99 (23:05) & 10.70 (26:20) \\ 
2006 07 06  & 08:13 & 1 & 2.69 & 1.20 & 0.25 & 139.59 & 911 & 3.32 (26:35) & 0.05 (26:25) & 0.08 (15:45) \\ 
2006 12 13  & 02:14 & 1 & 9.40 & 51.67 & 3.40 & 2245.36 & 1774 & 1100.00 (09:45) & 52.00 (05:35) & 25.00 (10:25) \\ 
\end{tabular}
\end{sidewaystable}

\begin{sidewaystable}
\footnotesize
\caption[]{Parameters of ICME events (see Table~\ref{Tab_SoWi}).} 
\vspace{1ex}
\label{Tab_ICME}
\begin{tabular}{lrrrrrrrrrr}
\hline
\hline
\multicolumn{8}{l}{Solar Activity} & \multicolumn{3}{l}{SEP peak intensity} \\
Date & Time & Qual & $\nu_{\rm max}$ & $\Phi_{\mu}$ & $I_{\rm SXR}$ & $\Phi_{\rm SXR}$ & $V_{\rm CME}$ & $J_{\rm e}(38 \; \rm keV)$ & $J_{\rm e}(175 \; \rm keV)$ & $J_{\rm p}(15 \; \rm MeV)$ \\
yyyy mm dd & hh:mm & & [GHz]    &              &           &              &           & ($\times 10^3$) & ($\times 10^3$) &  ($\times 1$) \\
(1) & (2) & (3) & (4) & (5) & (6) & (7) & (8) & (9) & (10) & (11) \\
\hline
1998 05 02  & 13:31 & 1 & 8.80 & 9.49 & 1.10 & 178.28 & 938 & 300.49 (16:15) & 9.63 (16:20) & 5.20 (15:25) \\ 
1998 05 06  & 07:58 & 1 & 8.80 & 4.42 & 2.70 & 683.21 & 1099 & 1550.00 (08:40) & 41.04 (08:25) & 8.70 (09:20) \\ 
1999 12 28  & 00:39 & 1 & 9.40 & 4.67 & 0.45 & 87.79 & 672 & 32.68 (05:00) & 1.20 (04:55) & 0.00 \\ 
2000 06 25  & 07:17 & 1 & 2.69 & 1.11 & 0.19 & 190.06 & 1617 & 0.00 & 0.00 & 0.05 (22:35)\\ 
2001 03 29  & 09:57 & 1 & 8.80 & 42.41 & 1.70 & 904.38 & 942 & 68.67 (21:55) & 3.44 (17:05) & 0.90 (26:50) \\ 
2001 04 02  & 21:32 & 2 & 15.40 & 44.84 & 20.00 & 7218.18 & 2505 & 729.09 (30:10) & 24.36 (28:50) & 10.63 (29:15) \\ 
2002 04 21  & 00:43 & 1 & 4.99 & 88.15 & 1.50 & 2747.99 & 2393 & 1230.00 (21:20) & 36.18 (13:25) & 81.50 (13:15) \\ 
2003 05 31  & 02:13 & 1 & 8.80 & 25.52 & 0.93 & 221.54 & 1835 & 547.00 (05:30) & 10.10 (05:00) & 0.70 (06:20) \\ 
2003 10 29  & 20:37 & 2 & 15.40 & 113.40 & 10.00 & 3563.80 & 2029 & 1780.00 (26:20) & 140.00 (26:25) & 65.20 (27:00) \\ 
2004 11 10  & 02:00 & 1 & 17.00 & 14.62 & 2.50 & 568.76 & 3387 & 168.70 (09:30) & 4.78 (09:30) & 10.62 (10:10)  \\ 
\hline
\end{tabular}
\end{sidewaystable}

\section{Observations and Data Analysis}
\label{Sec_Obs}

The data set considered for this study is based on SEP events associated with flares of soft X-ray classes M (peak flux $I_{\rm SXR}$ between $10^{-5}$ and $10^{-4}$~W~m$^{-2}$) and X ($I_{\rm SXR} \geq 10^{-4}$~W~m$^{-2}$) at western longitudes during the period 1997-2006. They are listed in \inlinecite{Can:al-10}. We used proton time profiles in the 15-40~MeV energy range observed by the GOES satellites, provided by the Ionising Particle ONERA DatabasE (IPODE; courtesy D.~Boscher) developed at the Office National d'Etudes et Recherches A\'erospatiales (ONERA) in Toulouse. It hosts data that were carefully compared between simultaneously observing GOES spacecraft and corrected for evident outliers. Near-relativistic electrons (energy ranges 38-53 and 175-315 keV) were observed by the {\it Electron, Proton, and Alpha Monitor} (EPAM) aboard the {\it Advanced Composition Explorer} (ACE) spacecraft \cite{Gol:al-98}\footnote{Level 2 data with 5~min integration from \url{http://www.srl.caltech.edu/ACE/ASC/level2/lvl2DATA_EPAM.html}}. We used the measurements of the magnetically deflected electrons to ensure there is no confusion with protons. Peak intensities of the electrons in both energy channels, henceforth referred to as $J_{\rm e}$(38 keV) and $J_{\rm e}$(175 keV), and of the protons, $J_{\rm p}$(15 MeV), were determined as the maximum of the intensity time profiles after subtraction of a pre-event background that we assumed constant. The background-subtracted intensity profiles were checked visually. They do not always show a simple well-defined peak. Several events have a time profile with a flat maximum and possible superposed fluctuations or a superposed energetic storm particle (ESP) event accelerated by a shock wave in interplanetary space. In these cases we estimated visually the time when a representative value of the peak intensity due to solar acceleration was reached. This procedure introduced some tolerable uncertainty in the intensity measurements, but made the determination of the start-to-peak fluences of the particles ambiguous. Therefore we consider only peak SEP intensities in the present study.

The association of a given SEP event with the parent eruptive activity in the solar corona was also based on \inlinecite{Can:al-10}. We examined flare positions and compared the time evolution at soft X-rays (SXR) and radio waves with the particle time profiles in order to eliminate cases where several flare-CME events could be associated with one SEP event. Following \inlinecite{Mit:al-13} two categories of events were identified according to the interplanetary magnetic field (IMF) configuration along which the SEPs propagated to the spacecraft: events where the SEPs were detected in the standard solar wind, henceforth referred to as SoWi events, and those where the spacecraft was within an interplanetary coronal mass ejection (ICME events). SEP events starting between one day before the onset and one day after the end of an ICME at Earth, as reported by \inlinecite{Ric:Can-10}, were discarded because the particle transport in the disturbed IMF could affect the SEP intensities. The selected events are listed in Tables~\ref{Tab_SoWi} (SoWi events) and \ref{Tab_ICME} (ICME events). The peak particle intensities are quoted in columns~9-11 together with the times when they were measured. These times refer to the date in column~1 and can therefore exceed 24~UT. When the SEP time profile did not allow a clear identification of the peak because the event did not emerge out of the background or because its time profile was complex, with fluctuations comparable to the peak intensity, the intensity was attributed the value 0, and was not used in the statistical evaluations.

For each SEP event the eruptive solar activity is characterised by parameters of the associated SXR and microwave (frequency range 1~GHz to some tens of GHz) bursts and by the speed $V_{\rm CME}$, projected onto the plane-of-sky, of the associated CME. SXR time histories observed by the GOES satellites\footnote{provided by NASA/GSFC at \url{http://umbra.nascom.nasa.gov/goes/fits/}} were used to determine the peak flux $I_{\rm SXR}$ and the start-to-peak fluence $\Phi_{\rm SXR}$ in the 0.1--0.8~nm channel. Both values were determined after subtraction of a pre-event background assumed constant. The values of $V_{\rm CME}$ were taken from linear fits to the time-height trajectory of the CME front as provided in the CME catalogue \cite{Yas:al-04}\footnote{NASA/GSFC and the Catholic University of America at \url{http://cdaw.gsfc.nasa.gov/CME_list/}} of SoHO/LASCO \cite{Bru:al-95}. The  microwave measurements were extracted from the data sets of the {\it Nobeyama Radio Polarimeters} (NoRP)\footnote{\url{http://solar.nro.nao.ac.jp/norp/html/event/}} and the {\it Radio Solar Telescope Network} (RSTN)\footnote{\url{http://www.ngdc.noaa.gov/stp/space-weather/solar-data/solar-features/solar-radio/rstn-1-second/}} of the US Air Force. The NoRP  \cite{Nak:al-85}, operated by the National Astronomical Observatory of Japan, measure whole-Sun flux density time histories at 1, 2, 3.75, 9.4, 17 and 35 GHz from about 23~UT to 07~UT. The RSTN measures whole Sun  flux densities at 0.24, 0.41, 0.61, 1.4, 2.7, 4.9, 8.8 and 15.4 GHz during 24 hours a day obtained from observatories in Sagamore Hill (Massachusetts), Palehua (Hawaii), Learmonth (Australia) and San Vito (Italy).

\begin{figure}
\centerline{
\includegraphics[width=0.5\textwidth]{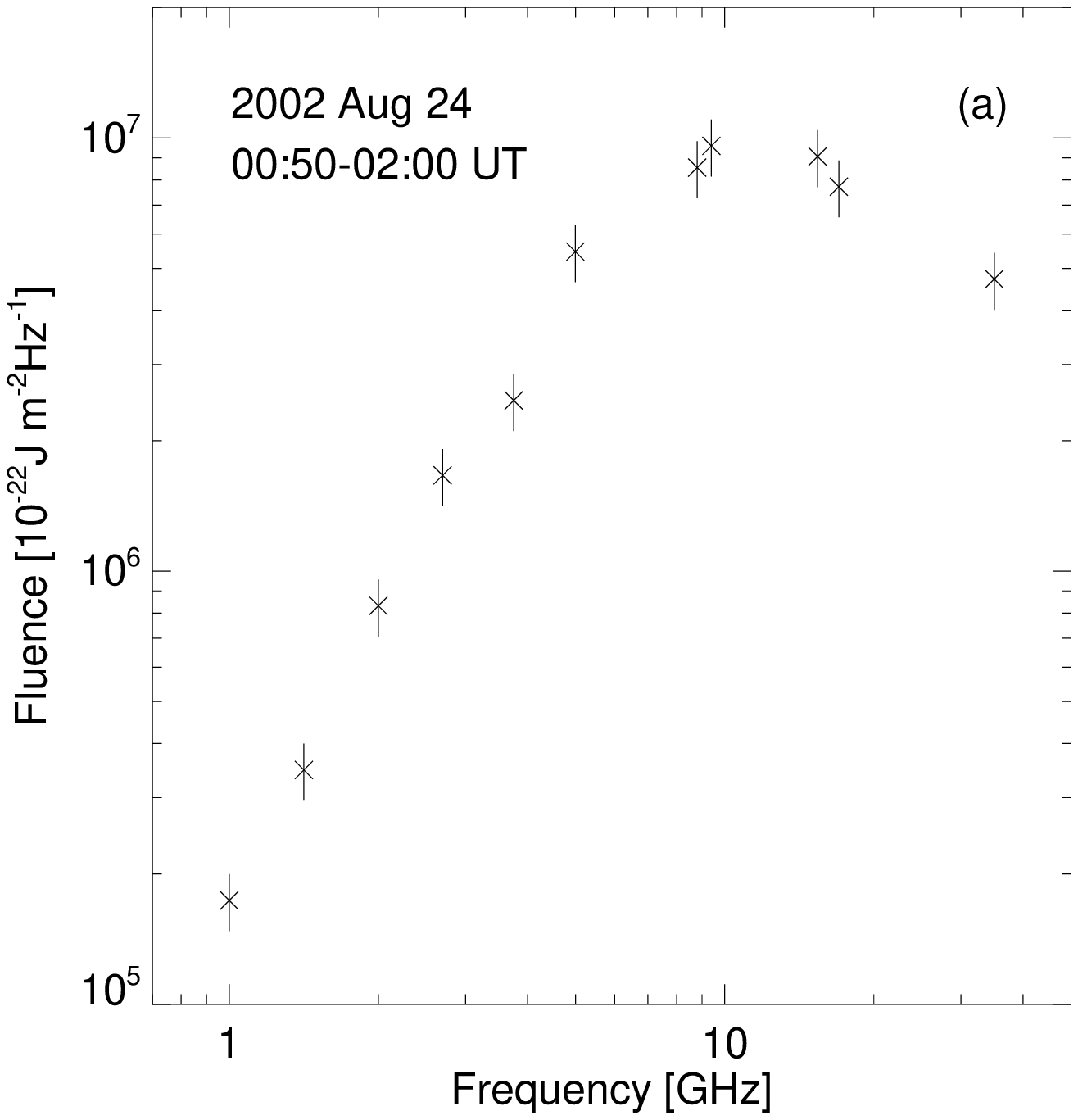}
\hspace*{0.01\textwidth}
\includegraphics[width=0.5\textwidth]{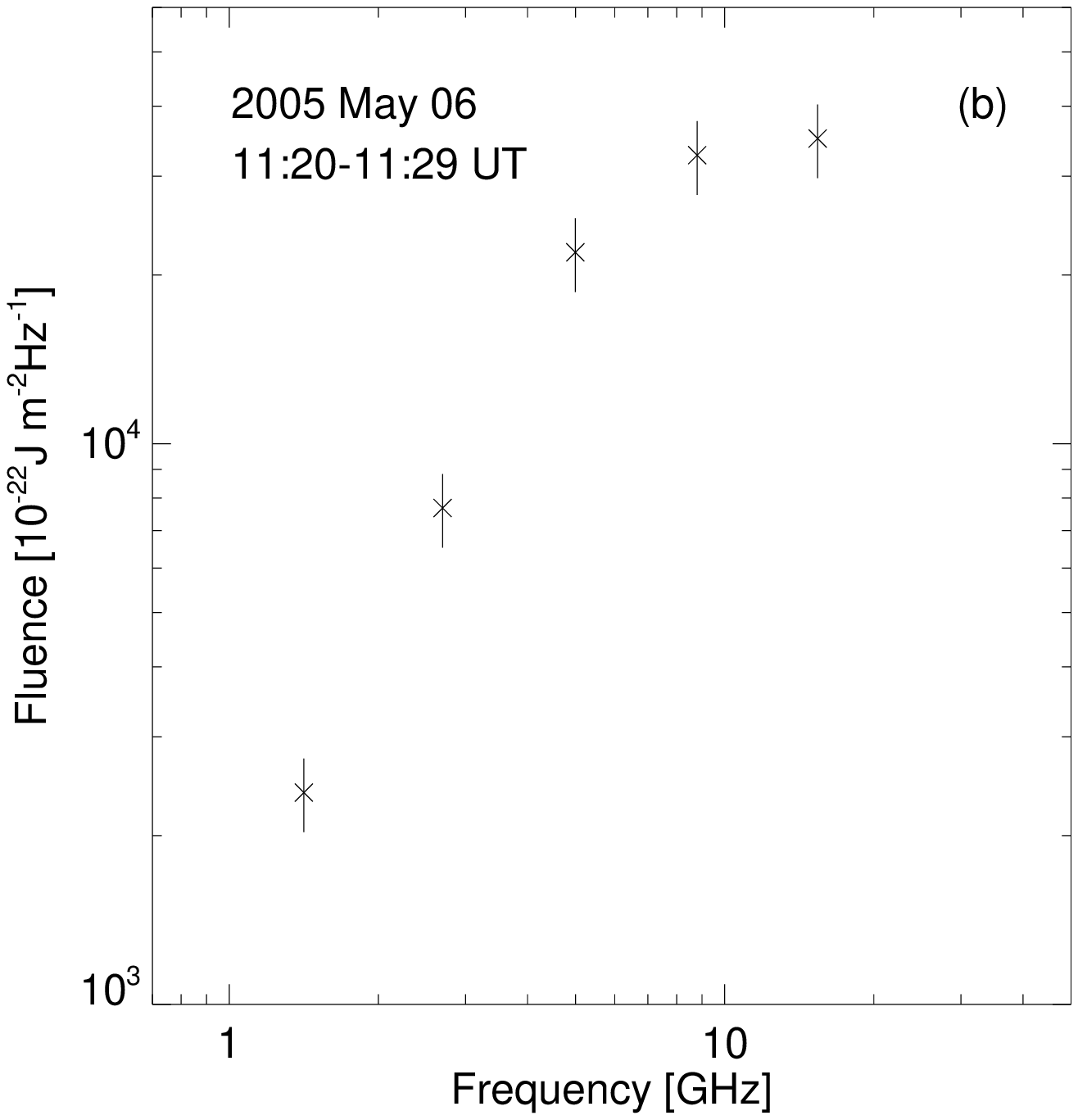}
}
\caption[]{The microwave fluence spectra of two events with different values of the quality flag. (a) An event with quality flag~1, indicating that the maximum of the fluence spectrum lay within the observed frequency range (combined data from NoRP and one RSTN station). (b) An event with a quality flag~2, where the maximum of the fluence spectrum occurred outside of, but close to the observed frequency range (data from the San Vito RSTN station). 
 }
\label{Fig_flsp}
\end{figure}

\subsection{Microwave Fluence}

The fluence was calculated at each frequency $\geq 1$~GHz during the entire burst. A constant background level was determined before or after the burst, and was subtracted. The peak value of the fluence spectrum, $\Phi_\mu$, was identified together with the frequency where it occurred, $\nu_{\rm max}$. In four weak events a thermal bremsstrahlung contribution was identified by its slowly evolving time profile and the frequency-independent flux density at high frequencies. The thermal and non thermal contributions could in all cases be clearly distinguished, and the thermal contribution was subtracted where necessary. Even in the case with the clearest bremsstrahlung component its contribution at the frequency of maximum microwave fluence did not exceed 10\%. The measured $\nu_{\rm max}$ and $\Phi_\mu$ are given in columns~4 and 5, respectively, of Tables~\ref{Tab_SoWi} and \ref{Tab_ICME}. 

In several cases the same burst was observed by more than one station. Figure~1(a) shows a fluence spectrum combining observations of one RSTN station and NoRP. The observations usually agreed reasonably well. In cases with multiple RSTN observations of equal quality at the same frequency the average fluence was taken. For some events the maximum microwave fluence lay outside the frequency range, notably when only RSTN observations were available. We assigned a quality flag to each  microwave burst, varying from 1 in cases where the peak fluence was clearly identified to 3 where it lay outside the observed frequency interval.  Quality flag 2 designates events where the fluence peak was not observed, but  the spectral shape showed that it was not far above the highest frequency observed. An example is shown in Figure~1(b). Events with quality flag 3 were not used in the  subsequent correlation studies, events with quality flag 2 were only employed to evaluate rank correlation. Tables \ref{Tab_SoWi} and \ref{Tab_ICME} actually contain only events where the microwave fluence maximum occurred within the observed frequency range or close to it. The sample analysed in the following comprises 34 SoWi events and 10 ICME events. The suffix `c' is added to the quality flag when the microwave emitting electrons did not escape from the corona, as will be discussed in the following.


\begin{figure}
\centerline{
\includegraphics[width=0.5\textwidth]{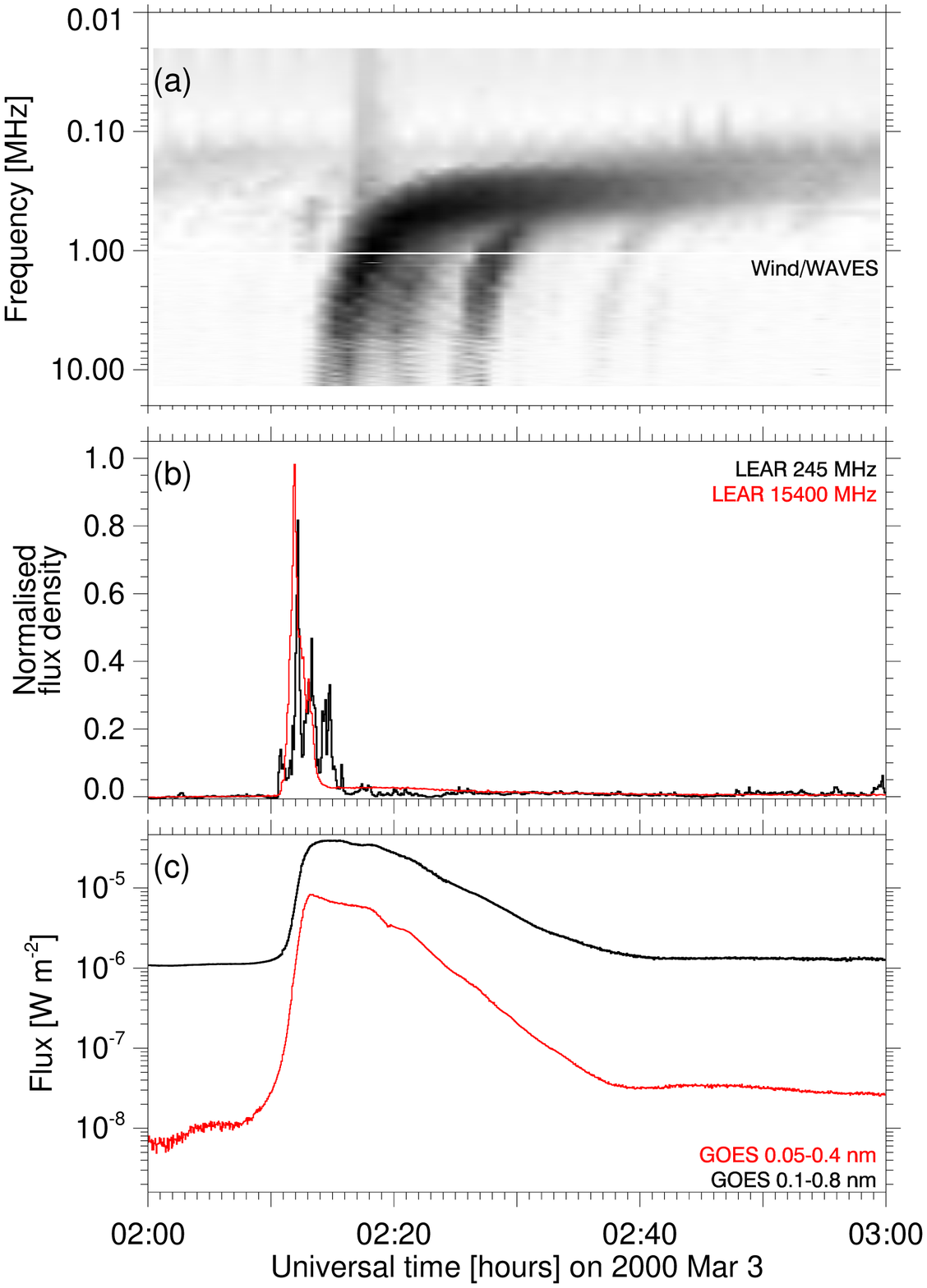}
\hspace*{-0.01\textwidth}
\includegraphics[width=0.5\textwidth]{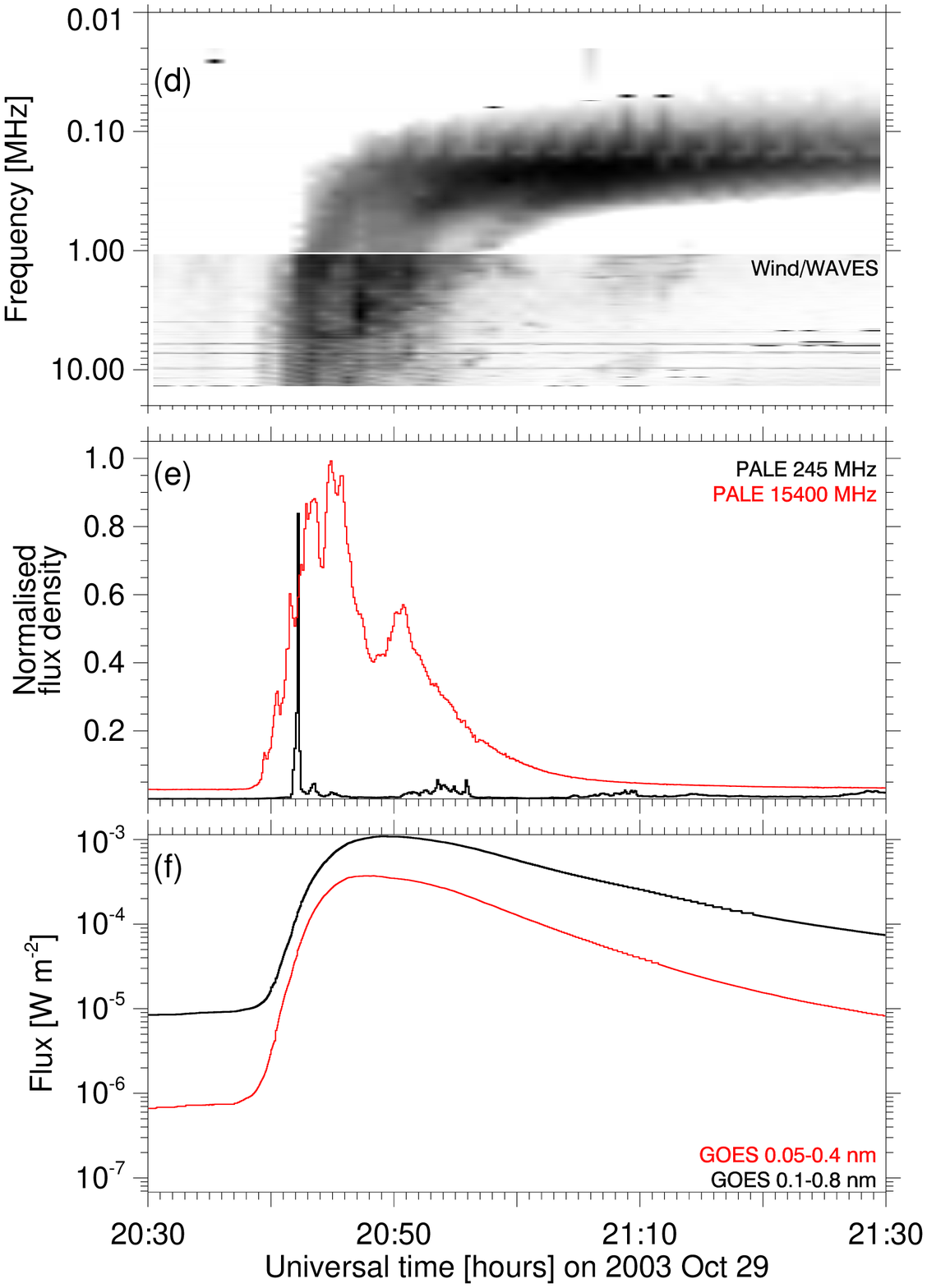}
\caption[]{Time histories of X-ray and radio emissions of a confined and an unconfined event, respectively. (a-c) 3 March 2000, (d-f) 29 October 2003.
From top to bottom: (a, d) decametre-to-hectometre (DH)wave emission ({\it Wind}/WAVES); (b, e) flux density time profiles at two frequencies in the microwave (15.4 GHz) and metre wave (245~MHz) range (RSTN Network); and 
(c, f) soft X-ray flux (GOES; black line 0.1 - 0.8 nm, lightly shaded line - red in the online version - 0.05 - 0.4 nm).}}
\label{Fig_conf}
\end{figure}

\subsection{Particle Escape from the Flare Site}
\label{Sec_Obs_esc}

Correlations between the parameters of the SEPs and of the associated flare have a physical sense only when the flare-accelerated particles actually escape from the parent active region. It was shown earlier \cite{Kle:al-10,Kle:al-11} that flare-accelerated electrons traced by their microwave emission may remain confined in the parent active region. In this case the microwave burst has no low-frequency counterpart. There is especially no decametric-to-hectometric type III emission (DH III). Type~III bursts are emitted by electron beams at the local electron plasma frequency or its harmonic (see  \citeauthor{Nin:al-08}, \citeyear{Nin:al-08}, and references therein). They are short bursts with a rapid drift from high to low frequencies that translates the propagation of the electron beams from the low to the high corona. This property makes them ideal tracers of the escape of energetic electrons through the high corona and the interplanetary medium. DH~III bursts are readily observable by the {\it Wind}/WAVES radio spectrograph \cite{Bou:al-95}. It is well known that they accompany SEP events \cite{Can:al-02,Can:al-10}. But only if these bursts occur during the microwave burst do they signal the escape of flare-accelerated electrons, and probably flare-accelerated particles in general, to interplanetary space. Type III bursts that occur after the microwave burst may signal electron beams accelerated at coronal shocks, as illustrated, for instance, in \inlinecite{Bou:al-98} and \citeauthor{Man:al-03} (\citeyear{Man:al-03}, their Figure~1). An example of a confined microwave burst that was followed by type~II and type~IV emission associated with DH~type~III bursts is shown in Figure~2 of \inlinecite{Kle:al-11}. In order to make sure that particles escape from the flaring active region during the microwave burst we therefore request that DH~type~III burst emission occur together with the microwave burst, allowing only for a delay of not more than a minute, which the type~III emitting electrons may need to reach the level where plasma emission in the {\it Wind}/WAVES frequency range can originate.

Following this reasoning, we considered the flare-accelerated particles to be confined when a significant part or all of the microwave emission occurred without a DH~III burst counterpart, while in microwave bursts with simultaneous type~III bursts flare-accelerated particles were likely able to escape from the corona. We refer to these bursts as confined and unconfined microwave events, respectively. Figure~2 displays examples of the two categories: the microwave burst at 15.4~GHz on 3 March 2000 (Figures 2(a)-(c)) had no accompanying DH~III burst during its entire rise phase. The type~III burst group started about a minute after the microwave peak and accompanied the decay of the microwave and SXR emission. This suggests that most of the flare-accelerated particles were confined in the corona, and could not contribute to the SEP event. The burst on 29 October 2003 (Figures 2(d)-(f)) was accompanied by DH type~III emission for its entire duration, and is therefore considered as a case where flare-accelerated particles contributed to the SEPs. Six microwave events of the 34 SoWi events of our sample were confined, and none of the ten ICME events. As said before, the confined microwave events  are indicated by adding the suffix `c'  to the quality flag in column~(3) of Tables~\ref{Tab_SoWi} and \ref{Tab_ICME}. In most of the confined events the DH~III bursts started a few minutes after the peak or end of the microwave burst, as in Figure~2. In the 13~July 2005 event it is the later phase of the microwave emission that is confined, as discussed in  \inlinecite{Mln:al-12}.

\section{Results}
\label{Sec_Sta}

For all parameters of the SEPs and the coronal activity discussed in this section both Pearson's correlation coefficient between the logarithms of the variables and the Spearman rank correlation coefficient were calculated. The former assumes a linear relationship between the logarithms of the two variables, whereas the latter is non parametric. We found identical results to within less than a standard deviation. This suggests that the linear relationship between the logarithms is a reasonable assumption, and so from now on only Pearson's correlation coefficient will be considered. Because of the small number of ICME events, we did not attempt a separate correlation study for them. All statistical correlations use SoWi events and ICME events at a time.

\subsection{Correlation between SEP Parameters}
\label{Sec_corr_SESE}

The coefficients of correlation between the peak intensities of different SEP species and energies are listed in Table~\ref{Tab_sep} for unconfined events. The number of confined events is too small to allow for a separate statistical analysis. In order to estimate a statistical uncertainty of the correlation coefficient, we used the bootstrap method \cite{Wal:Jen-12}: the correlation coefficient was calculated for $N$ couples of values chosen at random within the set of $N$ observations, and this was repeated $n=5000$ times. The mean value of the $n$ runs was adopted as the correlation coefficient, with the standard deviation as statistical uncertainty. The numbers of events $N$ used for the different correlations differ. Out of 33 events, which were unconfined and had well-determined microwave fluence (quality flag 1), four had no reliable measurement of the proton intensity, one no reliable measurement of the electron intensity. The SXR fluence could not be determined in one event. Our sample hence has a complete set of measurements for the deka-MeV protons in 29 SEP events and for the near-relativistic electrons in 32 events.

\begin{table}[h] 
\caption[]{Correlations between peak intensities of different SEP populations in unconfined microwave events.}
\center
\begin{tabular}{llll}
\hline
                          & $\log_{10}  J_{\rm e}$(175 keV) &  $\log_{10} J_{\rm p}$(15 MeV)\\
\hline
%
$\log_{10} J_{\rm e}$(38 keV)   & 0.95 $\pm$  0.01  & 0.79 $\pm$  0.08  \\
$\log_{10} J_{\rm e}$(175 keV) &  -                                              & 0.88 $\pm$  0.04 \\
\hline
\end{tabular}
\label{Tab_sep}
\end{table}

The SEP intensities display strong correlations, successively lower in the following order, but all highly significant: between the peak intensities of electrons, $J_{\rm e}$, in the two energy channels (Pearson correlation coefficient $\rho=0.95$), between the high-energy electrons and the deka-MeV protons, $J_{\rm p}$ ($\rho=0.88$), and finally between the low-energy electrons and $J_{\rm p}$ ($\rho=0.79$). The probability to get the same or a higher correlation coefficient from an unrelated sample is below 0.001\%. There is no difference within the uncertainties between the correlations of the entire event set and of the unconfined events alone. This may of course be due to the small number of confined events.  
 
\begin{table}[h] 
\caption[]{Correlations between variables of the solar activity in the entire event sample (quality flag 1, confined and unconfined events, with or without reliable SEP measurement).}
\footnotesize
\center
\begin{tabular}{lllll}
\hline
                       & $\log_{10} I_{\rm SXR}$  &  $\log_{10} \Phi_{\rm SXR}$ & $\log_{10} V_{\rm CME}$\\
\hline
 $\log_{10} \Phi_\mu$        & 0.65 $\pm$  0.09  & 0.84 $\pm$  0.03 & 0.65 $\pm$  0.09 \\
$\log_{10} I_{\rm SXR}$          &  -                           & 0.72 $\pm$  0.07 & 0.31 $\pm$  0.13  \\
$\log_{10} \Phi_{\rm SXR}$     &  -                           &  -                          & 0.61 $\pm$  0.10 \\

\hline
\end{tabular}
\label{Tab_ssi}
\end{table}

\subsection{Correlation between Solar Activity Parameters}
\label{Sec_corr_SoSo}

 \inlinecite{Kah-82b} introduced the term {\it Big Flare Syndrome} (BFS) to describe the empirical fact that there is a correlation between any two parameters measuring the magnitude of a flare event, independent of the detailed physical relationship between them. In order to assess if this feature affects the correlations with SEP parameters, we first consider the correlation between the parameters that characterise the solar activity associated with SEP events. Table~\ref{Tab_ssi} lists the correlation coefficients. Again the results for unconfined events and for the entire event sample are identical to within a standard deviation. Only the correlation for the entire event sample is shown. 

The strongest, highly significant correlation appears between the microwave fluence and the start-to-peak SXR fluence. Both fluences have similar correlations with the CME speed in the plane of the sky, with correlation coefficients that are lower than between the two fluences. The lowest correlation is found between the peak SXR flux and the CME speed. It is actually not significant, with a 5.7\% probability to get the same or a higher correlation coefficient for a set of 38 measurements of two unrelated parameters. To within statistical uncertainties this correlation coefficient is consistent with those obtained by \inlinecite{Vrs:al-05},  \inlinecite{Bei:al-12} and \inlinecite{Mit:al-13}, but the sample size in these works was bigger, and so was the significance of the correlation reported there.
 
 The pronounced correlations between different parameters describing the solar activity associated with SEP events will have to be considered when we discuss correlations between these activity parameters and SEP peak intensities.
%

\begin{figure}
\centerline{
\includegraphics[width=1\textwidth]{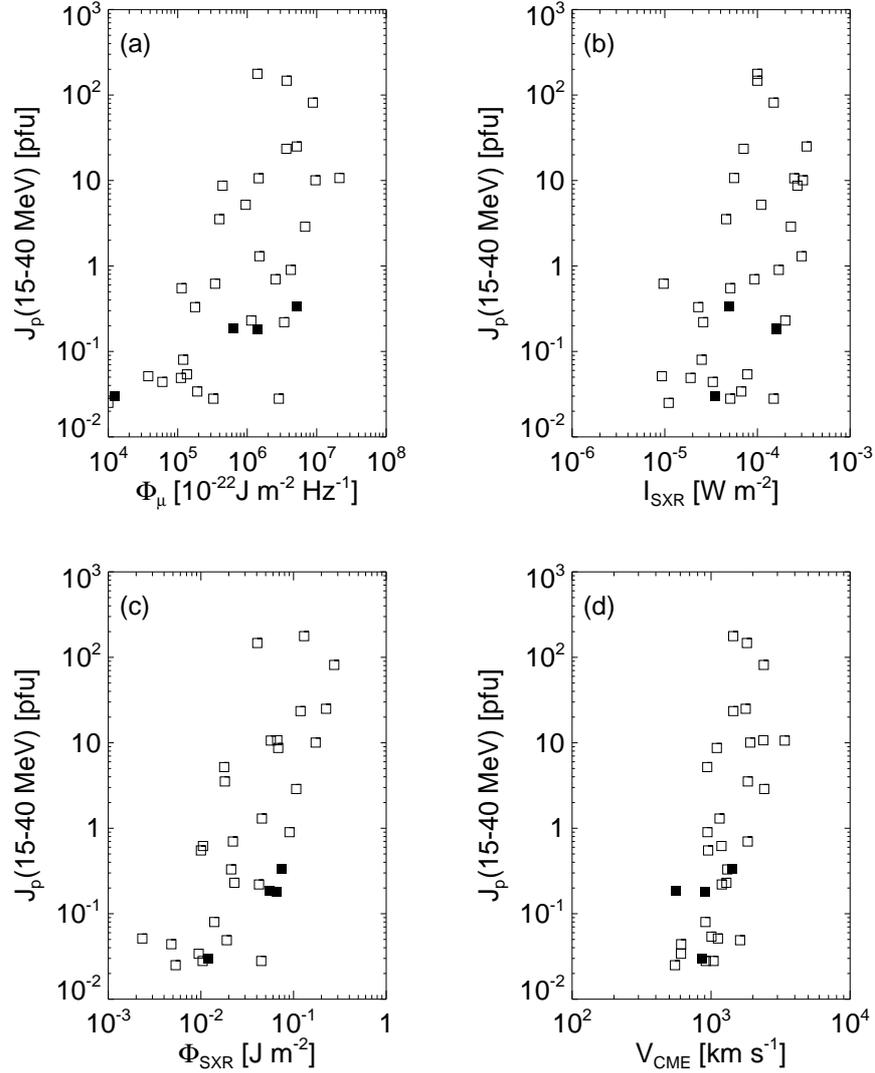}
\caption[]{Scatter (log-log) plots of proton peak intensity $J_{\rm p}$ versus microwave peak fluence $\Phi_{\mu}$, peak SXR flux $I_{\rm SXR}$, start-to-peak SXR fluence $\Phi_{\rm SXR}$ and CME speed $V_{\rm CME}$. Unconfined events are plotted by open squares, confined events by filled squares.}
}
\label{Fig_reg_p}
\end{figure}

\begin{figure}
\centerline{
\includegraphics[width=1\textwidth]{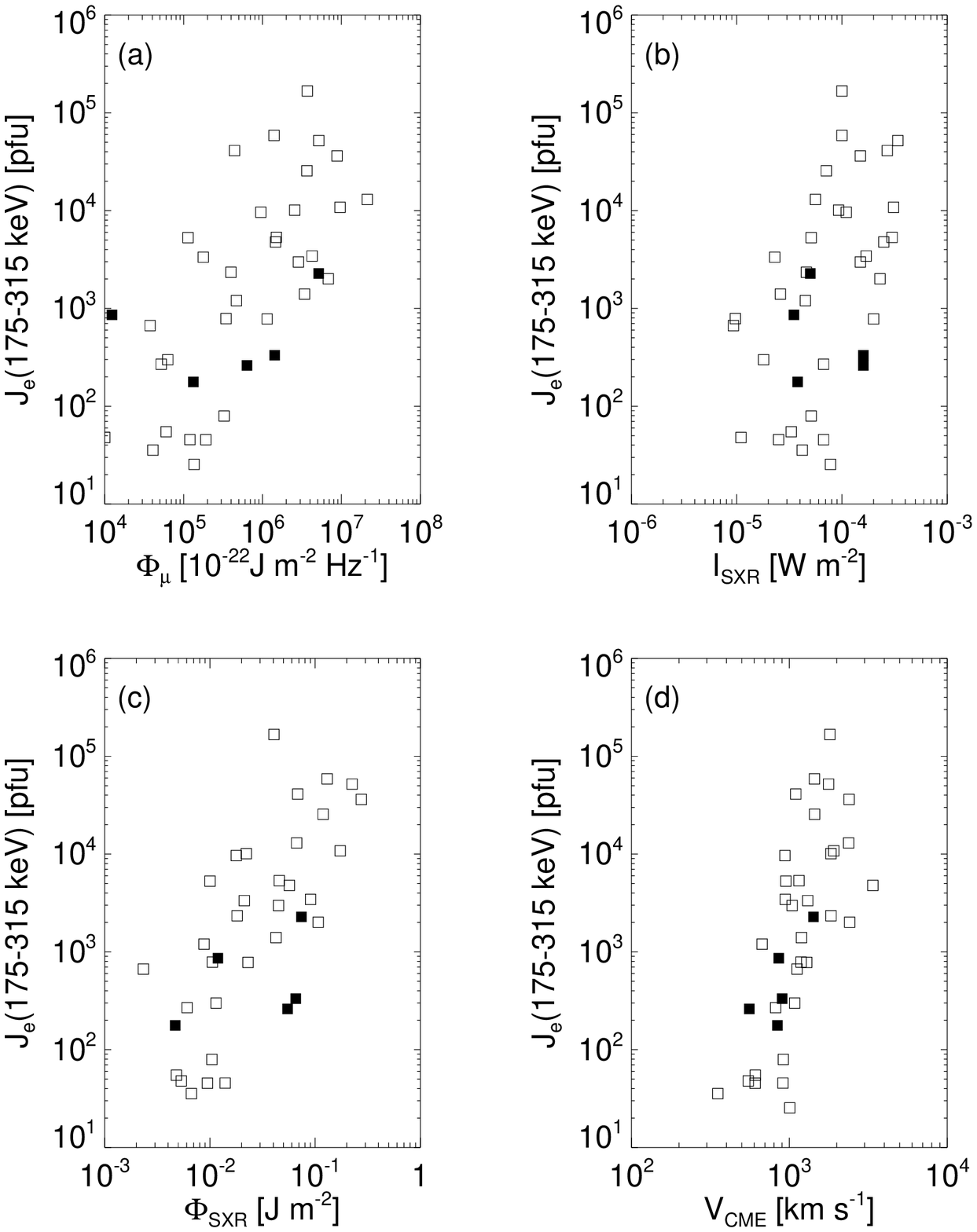}
\caption[]{Scatter (log-log) plots of the electron peak intensity $J_{\rm e}$(175 keV) versus microwave peak fluence $\Phi_{\mu}$, peak SXR flux $I_{\rm SXR}$, start-to-peak SXR fluence $\Phi_{\rm SXR}$ and CME speed $V_{\rm CME}$. Unconfined events are plotted by open squares, confined events by filled squares. 
}}
\label{Fig_reg_e}
\end{figure}

\subsection{Correlation between Parameters of Solar Activity and SEPs}
\label{Sec_corr_SoSE}

Scatter plots of the SEP peak intensities as functions of the different parameters of solar activity are displayed in Figures~3 and 4. The symbols distinguish unconfined (open squares) and confined (filled squares) microwave events. Only four confined events enter into the statistics for the protons, one more into that of the electrons. All are situated in the lower two of the four orders of magnitude spanned by the SEP intensities. The plots show the usual noisy correlation between peak SEP intensities and parameters of solar activity, with no obvious difference between deka-MeV protons (Figure~3) and near-relativistic electrons (Figure~4).

\begin{table}[h]     
\footnotesize
\center
\begin{tabular}{llll}
\hline
                        & $\log_{10} J_{\rm e}$(38 keV) &  $\log_{10} J_{\rm e}$(175 keV) &  $\log_{10} J_{\rm p}$(15 MeV)\\
                             \hline
\multicolumn{4}{c}{Pearson's correlation coedfficients:} \\
$\log_{10} \Phi_\mu$      & 0.61 $\pm$  0.09  & 0.72 $\pm$  0.07  & 0.67 $\pm$  0.09  \\
$\log_{10} I_{\rm SXR}$       & 0.35 $\pm$  0.11   & 0.53 $\pm$  0.09  & 0.54 $\pm$  0.10  \\
$\log_{10} \Phi_{\rm SXR}$  & 0.65 $\pm$  0.08   & 0.75 $\pm$  0.06  & 0.76 $\pm$  0.08  \\
$\log_{10} V_{\rm CME}$     & 0.65 $\pm$  0.09   & 0.68 $\pm$  0.08  & 0.67 $\pm$  0.08  \\
%
\multicolumn{4}{c}{Partial correlation coefficients:} \\
%
$\log_{10} \Phi_\mu$      & -0.03 $\pm$ 0.17 & 0.05 $\pm$ 0.19  & -0.10 $\pm$ 0.22  \\
$\log_{10} I_{\rm SXR}$       & -0.14 $\pm$  0.16 & 0.10 $\pm$ 0.16  &  0.06 $\pm$ 0.17  \\
$\log_{10} \Phi_{\rm SXR}$  &  0.31 $\pm$  0.16 & 0.27 $\pm$ 0.18  &  0.42 $\pm$ 0.20  \\
$\log_{10} V_{\rm CME}$     &  0.35 $\pm$  0.14 & 0.34 $\pm$ 0.16  &  0.36 $\pm$ 0.18  \\
\hline
\end{tabular}
\caption[]{Correlations between parameters of the solar activity and SEP peak intensities (unconfined events).}
\label{Tab_sosei}
\end{table}

The correlation coefficients between SEP peak intensities and the parameters of the solar activity are listed in Table~\ref{Tab_sosei}. The Pearson correlation coefficients in the upper table show that to within the statistical uncertainty of one standard deviation the correlations with CME speed, SXR fluence and microwave fluence are the same. The correlations between SEP peak intensities and SXR peak flux are much weaker -- the probability to get the same or a higher correlation coefficient from unrelated data sets is 4.8\%. 

\subsection{Partial Correlations}
\label{Sec_corr_part}

Given the strong interdependence between the parameters used to describe the eruptive solar activity associated with the SEP events (Section~\ref{Sec_corr_SoSo}), the interpretation of the correlations presented in Section~\ref{Sec_corr_SoSE} is unclear. The peak SEP intensity may be related to all quantities with a significant Pearson correlation coefficient, but correlations may also appear because of the interdependence of the solar parameters. In order to investigate if one or several of the parameters describing the solar activity are more strongly correlated with SEP peak intensity than others, we present here calculations of the partial correlation between the different parameters. 

Partial correlations consider the relationship between two variables $a$ and $b$ after removal of the relationship with the other variables $c, d, \ldots $, called control variables. More specifically, one considers two variables out of a larger set, for instance the peak SEP intensity and the SXR fluence, while the CME speed, the microwave fluence and the SXR peak flux are control variables. For both variables the residuals are calculated, {\it i.e.} the difference between the original quantity and the linear regression with the control variables. The partial correlation coefficient $\rho_{a,b | c, d, \ldots}$ is the correlation coefficient of the residuals: it quantifies the linear dependence between variables $a$ and $b$, with the difference that the influence of the control variables has been eliminated. We can thus expect that $|\rho_{a,b | c, d, \ldots}| \le |\rho_{a,b}|$, where  $\rho_{a,b}$ is the classical ({\it i.e.} Pearson) correlation coefficient. The partial correlation coefficients between the SEP peak intensities and all considered parameters of the solar activity  are listed in the lower part of Table~\ref{Tab_sosei}, together with the statistical uncertainties from the bootstrap method. This table shows the following:
\begin{itemize}
\item The partial correlations between the different parameters are indeed lower than the Pearson correlations in the upper part of the table. 
\item The event sample turns out to be rather small, implying large uncertainties of the partial correlation coefficients. Nevertheless it is clear that the only quantities with a significant relationship to the SEP peak intensities are the CME speed and the SXR start-to-peak fluence. The correlation coefficients for the three SEP types and the two parameters of solar activity are identical to within the statistical uncertainties.
\item Microwave peak fluence and SXR peak flux play no independent part. The significant Pearson correlation between SEP peak intensity and microwave peak fluence found in Section~\ref{Sec_corr_SoSE} is thus apparently due to the strong partial correlation of the microwave fluence with SXR fluence (partial correlation coefficient $0.52 \pm 0.13$) and CME speed (partial correlation coefficient $0.37 \pm 0.16$).
\end{itemize}

\begin{table}
\begin{tabular}{lrrr}
\hline
\hline
 & $\log_{10} J_{\rm e}\left(38 \; \rm keV \right)$ & $\log_{10} J_{\rm e} \left(175 \; \rm keV \right)$ & $\log_{10} J_{\rm p}\left(15 \; \rm MeV \right)$ \\
(1) & (2) & (3) & (4) \\
\hline
$A$ & -0.17 $\pm$ 2.36 & -0.72 $\pm$ 2.21 &   -4.27 $\pm$ 2.62 \\
$B$ &  1.96 $\pm$ 0.70 & 1.77 $\pm$ 0.66 &  1.93 $\pm$ 0.75 \\
$C$ & 0.56 $\pm$ 0.20 & 0.90 $\pm$ 0.18 & 1.15 $\pm$ 0.24 \\
\hline
\hline
\end{tabular}
\caption[]{Total least squares fit to the SEP intensity in terms of CME speed and SXR fluence (unconfined events; Equation.~\ref{Eq_params}).}
\label{Tab_regr}
\end{table}

The linear regression with the two significantly correlated parameters, obtained from total least squares regression \cite{Gol:vLo-13}, reads
\begin{equation}
\log_{10} J_{\rm SEP} = A + B \log_{10} V_{\rm CME}  + C \log_{10} \Phi_{\rm SXR} \;
\label{Eq_params}
\end{equation}
where $V_{\rm CME}$ is in km s$^{-1}$, $\Phi_{\rm SXR}$ in J m$^{-2}$ and the constants are given in Table~\ref{Tab_regr}.

\section{Discussion}
\label{Sec_Disc}

The present work analysed statistically the relationship between SEP intensities and parameters of eruptive solar activity, using a sample of 44 SEP events between 1997 and 2006. By selection the SEP events were associated with strong flares (M and X class) in the western solar hemisphere. The associated CMEs had speeds between 350 and 3400~km~s$^{-1}$.  In the statistical analysis near-relativistic electrons in two energy ranges (38-53~keV and 175-315~keV) and deka-MeV protons (15-40~MeV) were considered, as well as parameters describing the different aspects of the associated eruptive solar activity: CME speed, flux  and fluence of the SXR burst, fluence of the microwave burst. Besides using several parameters describing solar activity we also considered if flare-accelerated electrons actually escaped from the corona. We used the presence of type~III bursts at decameter and longer waves as indicators of electron escape along open magnetic field lines. 

\subsection{Summary of Observational Results}

The observational results of this study are summarised as follows:
\begin{itemize}
\item In the vast majority (38/44) of the SEP events radio emission shows that flare-accelerated electrons had direct access to interplanetary space.
\item There is a strong correlation between the peak intensities of near-relativistic electrons and deka-MeV protons, with a slight (marginally significant) trend to be higher for the higher electron energy channel (Section~\ref{Sec_corr_SESE}).
\item A highly significant correlation was found between the peak intensities of near-relativistic electrons and deka-MeV protons with the projected CME speed, the maximum microwave fluence ({\it i.e.} the highest value of the fluence spectrum at frequencies above 1~GHz) and the start-to-peak SXR fluence (Section~\ref{Sec_corr_SoSE}). The correlation coefficients range between 0.61 and 0.76. Correlations with the three solar parameters were found identical to within less than one standard deviation. 
\item The correlation with the soft X-ray peak flux was found to be significantly lower for all SEP intensities considered, with correlation coefficients between 0.35 and 0.54.
\item Correlations exist also between the different parameters describing the eruptive solar activity (Section~\ref{Sec_corr_SoSo}): the strongest correlation, $\rho \simeq 0.8$, is found between the peak fluence of microwaves and the start-to-peak fluence of SXRs.
\item The partial correlation analysis (Section~\ref{Sec_corr_part}) shows that the interdependence of the different variables describing the eruptive solar activity is responsible  for some of the classical correlations. The only activity parameters that show some correlation with peak SEP intensity are the CME speed and the start-to-peak SXR fluence. The correlation between peak SEP intensity and both SXR peak flux and microwave peak fluence is spurious.
\end{itemize}

\subsection{Impact on the Interpretation of the Origin of SEPs}

\subsubsection{Correlations between the Intensities of Electrons and Protons}

The correlations between the peak intensities of deka-MeV protons and near-relativistic electrons are found to be strong. We exclude an observational artifact through a contamination of the electron channels by protons, because the electron intensities came from the deflected electron channels of ACE/EPAM. Correlation coefficients $\rho \simeq 0.9$ were reported \cite{Dbg:al-89} between the fluences of energetic electrons (0.07 MeV, 0.5 MeV) and protons above 25~MeV. \inlinecite{Pos-07} compared the initial rise of near-relativistic electrons and deka-MeV protons, and showed that the e-folding rise times of the two populations were closely related. The intensity correlation found in the present study is hence in line with earlier reports. These observations suggest that electrons and protons detected in space have a common history of acceleration and transport. 

\subsubsection{Escaping {\it vs.} Confined Flare-accelerated Particles} 

All SEP events considered here are accompanied by electron acceleration  in the flaring active region to near-relativistic energies, revealed by their gyro-synchrotron emission. The simultaneous presence of type~III bursts shows that a fraction of the flare-accelerated electrons has in most cases access to the interplanetary space along open magnetic field lines. We expect that this is also a valid conclusion for flare-accelerated protons, although we have no observational proof. The reason is twofold: first, because of the correlation between electron and proton intensities discussed above. Second, judging from the type III bursts, the electron acceleration region is either magnetically connected to open field lines in the high corona ({\it cf}. the scenario of \citeauthor{Msn:al-12a}, \citeyear{Msn:al-12a}), or else the electrons can scatter onto neighbouring open field lines. Unless accelerated at a remote place, the protons should therefore also be able to get access to open field lines. So we conclude that protons accelerated in the flaring active region likely have access to interplanetary space in the majority of the SEP events considered here. Those SEP events where observations point to the confinement of flare-accelerated particles, and which need a different accelerator, had peak intensities of at least an order of magnitude below the strongest SEP events of our sample. 

\subsubsection{Correlations between the Quantities Characterising Solar Activity}

One of the problems with the interpretation of statistical relationships between solar activity and SEPs is the interdependence of the different quantities used to characterise the solar activity. The notion of the {\it Big Flare Syndrome} \cite{Kah-82b} expresses this and emphasises the need to be cautious when translating correlations into physical relationships.

It is now well known that the acceleration of CMEs is closely related in time with the evolution of thermal energy release in the associated flare \cite{Zha:al-04,Bei:al-12}, suggesting a relationship between the CME speed and the peak flux or fluence of the SXR burst. A close relationship between thermal and non thermal energy release during a flare was also revealed earlier by the similarity of the SXR flux time profile during the rise phase of the burst and the fluence of the microwave \cite{Neu-68} and hard X-ray burst \cite{Den:Zar-93} -- a phenomenon known as the Neupert effect. \inlinecite{Kru:Ben-00} demonstrated a close relationship between the peak fluxes or peak luminosities in microwaves and SXRs. These interdependencies are clearly present in the activity accompanying the SEP events studied here.

\subsubsection{Correlation between SEP Intensities and Solar Activity}

Our study finds a distinct difference between the SEP correlations with different quantities describing the eruptive solar activity. The SEP peak intensities correlate significantly better with integral quantities (fluences) and CME speed than with SXR peak flux. The result in the upper Table~\ref{Tab_sosei} is consistent with the correlations between proton peak intensity above 10~MeV and SXR peak flux in \inlinecite{Cli:al-12} and between proton intensity and SXR fluence in \inlinecite{Cli:Die-13}. 
 \inlinecite{Kah-82b} had shown the preferred correlation with integral flare parameters before, comparing the correlation with SXR peak flux and microwave fluence at 9 and 15~GHz. He did not consider SXR fluence, however, and ascribed the difference of correlations with SXR and microwaves to the fact that the SXR emission is a thermal phenomenon, while the microwaves are of non thermal origin. This does not seem to be the correct interpretation, however, because the SXR fluence performs similarly to the microwave fluence in the pairwise classical correlation.

The basic problem with translating these correlations into the identification of physical relationships is the strong mutual correlation of the variables used to characterise the eruptive solar activity. The correlation of a given variable with an SEP peak intensity can therefore be ascribed to a direct physical relationship between the two, but also to a relationship of both variables with a third one.  The study of partial correlations, which account for the interdependence of the solar parameters, sheds new light on the problem. It appears clearly that among the solar activity parameters considered only the CME speed and the SXR start-to-peak fluence are significantly correlated with SEP peak intensity. To within the statistical uncertainties of the partial correlation coefficients the relationship with the two solar parameters is equally significant. The correlations with the three SEP categories -- 15-40~MeV protons, electrons in the 38-53~keV and the 175-315~keV range -- are also indistinguishable. This is a statistical confirmation of the idea  that both flare processes  and CME-driven shock waves contribute to the acceleration of deka-MeV protons and near-relativistic electrons in large SEP events, provided the flare-accelerated particles escape to interplanetary space and the flare is magnetically connected to the particle detector. The fact that the partial correlation is significant with the SXR fluence, but not with the SXR peak flux, means    that once the dependence of the SEP intensity on the SXR fluence is accounted for, the SXR peak flux has no more influence. This result is consistent with the finding \cite{Gar-04} that the correlation between proton peak intensity and SXRs is improved when more parameters than the mere SXR peak flux are included, among them the duration.

An intriguing result is that the microwave fluence has no statistically significant relationship with SEP intensity, unlike the SXR fluence. This is the case for deka-MeV protons as well as for near-relativistic electrons. Since the microwave emission is also emitted by near-relativistic electrons, such a difference was not expected, and for the time being we know of no convincing explanation. The microwave fluence does not only depend on the energetic electron spectrum, but also on the magnetic field strength and the magnetic field configuration. \inlinecite{Dbg:al-89} argued that the influence of the magnetic field could be removed when the microwave fluence is normalised by the square of the peak frequency, because the peak frequency grows with increasing magnetic field strength, too. We considered the correlations for this normalised fluence, too, but found them still lower than when the fluence itself was used. 

\subsubsection{Influence of Other Quantities on the SEP Peak Intensity}

Besides the CME speed and the SXR fluence, which describe in some sense the strength of the likely accelerators, quantities related to the conditions of particle propagation and to acceleration in the high corona and interplanetary space affect SEP intensities, but were not considered here. One quantity is the connection distance between the solar flare and the interplanetary field line through the particle detector. There is an apparent contradiction between the findings from statistical studies and multi-spacecraft observations of SEP events: the latter find a clear variation of peak SEP intensity with connection distance \cite{Kal-93,Lar:al-13,Ric:al-14,Drs:al-14}, while statistical analyses using single spacecraft observations reveal no effect \cite{Mit:al-13,Drc:al-14}. This is likely due to a bias of these statistical analyses: the distribution of peak intensities with connection distance or flare longitude broadens with increasing flare importance, as measured by the SXR peak flux, so that SEP events with larger connection distance are on average associated with stronger SXR bursts. This is clearly visible in Figure~4 of  \inlinecite{Mit:al-13}. It means that the SEP events observed at large longitudinal distance from the well-connected interplanetary field line are intrinsically stronger than those close to the field line, and that weaker events, which would have been observable near the interplanetary field line, are below the detection threshold at the larger distance. The uncertain connection distance of a single spacecraft hence contributes to blurring the relationship between SEP peak intensity and the parent eruptive activity. 

Other known or suspected effects on the SEP intensity are the influence of a supra thermal seed population on the shock-accelerated SEPs and the effects of the interplanetary magnetic field structure in which the SEPs propagate to the Earth. \inlinecite{Kah-01} examined the influence of the SEP energy spectra and the pre-event levels of energetic particles on the SEP peak intensity. He concluded that the pre-event level played a role, which could be interpreted as evidence for a supra thermal seed population that made local shock acceleration more efficient. Other authors \cite{Gop:al-04,Dng:al-13} invoked the possible role of multiple CMEs in enhancing the flux and energy of the SEPs. The claim that this reveals the role of CME interaction in SEP events has, however, been called into question   \cite{Kle-06,Kah:Vrl-14}. The effect of the interplanetary magnetic field structure was analysed by \inlinecite{Mit:al-13}. They showed that the correlation between peak SEP intensity and peak SXR flux is much higher when the SEPs are observed within an ICME than within the standard solar wind. They tentatively attributed this to a better magnetic connection between the solar activity and the Earth in the case of an ICME event (see their Figure~4, especially the top row). We have not enough events to conduct a separate study of SEPs in ICMEs and in the standard solar wind. 

\subsection{Conclusion: A Mixed Flare-CME Origin of deka-MeV Protons and Near-relativistic Electrons in SEP Events}

It is hence clear that relating peak SEP intensities in space to coronal activity involves many parameters. The
 research presented here yields a new type of statistical evidence for the mixed flare-CME contribution to SEPs suspected earlier \cite{Kal-03}. This is physically plausible: the only means to avoid a flare contribution to SEP events is to confine the flare-accelerated particles in coronal magnetic structures. It seems that the number of electrons in space is indeed only a very small fraction of the numbers required for the hard X-ray \cite{Kru:al-07} or gamma-ray \cite{Ram:al-93} emission, while the number of protons in space may be smaller or larger than the number required for nuclear gamma-rays at the Sun \cite{Ram:al-93}. But the vast majority of the SEP events of our sample show evidence that flare-accelerated particles escape to interplanetary space, probably along pre-existing open magnetic field lines. The few cases where the flare-accelerated particles appear to be confined have rather low SEP intensities. The evidence of a mixed flare-CME contribution pertains to protons of a few tens of MeV and to near-relativistic electrons. \inlinecite{Drc:al-14} showed that the correlation between SEP intensities and SXR peak flux increases with increasing particle energy, while the correlation with CME speed decreases. The correlation appears more pronounced with CME speed than with SXR peak flux at low energies, while the converse is true at high energies. The transition occurs in the 10-20 MeV range, but the error bars are large. This trend is consistent with scenarios with energy-dependent contributions of the two candidate processes, where CME shock acceleration dominates at the lower energies, while flare-related acceleration dominates at high energies \cite{Kle:Tro-01,Can:al-02,Can:Lar-06}.

\begin{acks}
The authors acknowledge D.~Boscher (ONERA Toulouse) for making the IPODE database of GOES particle measurements available to us. We acknowledge the generous supply of data from the ACE/EPAM particle instrument, the GOES particle and soft X-ray detectors, the {\it Wind}/WAVES radio spectrograph, the RSTN and NoRP radio instruments, and the {\it Radio Monitoring} web site \url{http://secchirh.obspm.fr/index.php} at Paris Observatory. Extensive use was made of the CME catalogue generated and maintained at the CDAW Data Center by NASA and The Catholic University of America in cooperation with the Naval Research Laboratory. SOHO is a project of international cooperation between ESA and NASA. The work presented here benefitted from partial financial support and from scientific cooperation within the SEPServer (Grant Agreement No. 262773) and HESPE (Grant Agreement No. 263086) projects of the {\it 7th Framework} programme of the European Union. This research was carried out within a collaboration between Egypt and France funded through  the IMHOTEP programme (contracts 23190YB and 27471UK). We are grateful to the Egyptian coordinator, Dr. M.~Shaltout, for his support. We also acknowledge support by the {\it Centre National d'Etudes Spatiales} (CNES). The referee and the editor are  thanked for their careful reading of the manuscript and helpful comments.
\end{acks}


\end{article}
\end{document}